\documentstyle[preprint,aps]{revtex}
\begin{document}
\draft
\preprint{
\parbox{4cm}{
\baselineskip=12pt
TMUP-HEL-9608\\
July, 1996\\
\hspace*{1cm}
}}
\title{An analysis on a supersymmetric chiral gauge theory\\
       with no flat direction}
\author{Noriaki Kitazawa\thanks{e-mail: kitazawa@phys.metro-u.ac.jp}}
\address{Department of Physics, Tokyo Metropolitan University,\\
         Hachioji-shi, Tokyo 192-03, Japan}
%\date{\today}
\maketitle
\begin{abstract}
The low energy effective theory
 of the $N=1$ supersymmetric $SU(5)$ gauge theory
 with chiral superfields in the $5^*$ and $10$ representations is constructed.
Instead of postulating the confinement of $SU(5)$ (confining picture),
 only the confinement of its subgroup $SU(4)$ is postulated (Higgs picture),
 and the effective fields are $SU(4)$-singlet but $SU(5)$-variant.
The classical scalar potential
 which ensures unique supersymmetric vacuum at the classical level
 is incorporated into the K\"ahler potential of the effective fields.
We show that
 supersymmetry and all other global symmetry are spontaneously broken.
The scales of these symmetry breaking
 and the particle spectrum including Nambu-Goldstone particles
 are explicitly calculated,
 and no large scale hierarchy is found.
\end{abstract}
%\pacs{}
\newpage

\section{Introduction}
\label{sec:intro}

It is expected that
 there are some unknown dynamics of the gauge theory
 by which the problems in the standard model are solved.
For instance,
 the natural scale hierarchy is expected in some class of
 chiral gauge theories (``tumbling gauge theories'')\cite{tumbling},
 and the mass hierarchy of quarks and leptons
 may be explained by virtue of such dynamics.
If the low-energy supersymmetry
 which may solve the naturalness problem exists,
 it may be spontaneously broken by the dynamics of the gauge theory.
But,
 since the non-perturbative effect of the gauge theory is hard to evaluate,
 our efforts for the concrete model building have been limited.

The method
 proposed by Seiberg et al.\cite{seiberg1,seiberg2,seiberg3} is remarkable,
 because it has a strong power to exactly determine
 the superpotentials of $N=1$ supersymmetric gauge theories.
Application of the method to various gauge theories
 will give us further new knowledge
 in addition to which has already been discovered
 on the supersymmetric QCD and so on
 \cite{seiberg1,seiberg2,seiberg3,others1,others2,others3}.
Especially,
 the gauge theories with no flat direction in the scalar potential
 should be extensively studied to understand real physics,
 since they have unique vacuum
 where the supersymmetry may be spontaneously broken.

The $N=1$ supersymmetric $SU(5)$ gauge theory
 with chiral superfields in the $5^*$ and $10$ representations
 has been extensively studied.
It is well known that
 the classical scalar potential of this theory has no flat direction,
 and ensures unique ground state with unbroken gauge symmetry.
The spontaneous breaking of supersymmetry
 is suggested by the explicit instanton calculation
 \cite{instanton1,instanton2}
 and the argument of the low energy effective theory\cite{effective}.
In the effective theory of ref.\cite{effective} two effective fields,
 $S \sim {\rm tr}(W^2)$ and $\Phi_A \sim \Phi^T \Omega W^2 \Phi$
 are introduced postulating the confinement of $SU(5)$,
 where $W$, $\Phi$ and $\Omega$ denote the chiral superfields of
 the $SU(5)$ gauge field strength,
 matters in the $5^*$ and $10$ representations, respectively.
It is argued that the supersymmetry is spontaneously broken
 by assuming a special K\"ahler potential
 and the general effective superpotential
 as far as the symmetry of the theory allows. 
In this picture
 all information on the dynamics is contained in the K\"ahler potential
 which can not be determined by the argument of the symmetry.

In this paper
 we construct an effective theory from the different starting point.
We do not postulate the confinement of $SU(5)$, but its subgroup $SU(4)$.
Namely, we take Higgs picture instead of confining picture
 (complementarity\cite{complementarity1,complementarity2})
 expecting the gauge symmetry breaking of $SU(5) \rightarrow SU(4)$.
This pattern of gauge symmetry breaking
 is suggested in the non-supersymmetric version of this theory:
 $SU(5)$ gauge theory with chiral fermions
 in the $5^*$ and $10$ representations\cite{tumbling}.
In this non-supersymmetric theory
 the most attractive channel hypothesis suggests
 a fermion pair condensation of the channel $10 \times 10 \rightarrow 5^*$
 which causes the breaking of $SU(5) \rightarrow SU(4)$.
Below its condensation scale,
 $SU(4)$ gauge theory with chiral fermion
 in the $4$, $4^*$ and $1$ representations
 is considered as the effective theory,
 and the subsequent QCD-like condensation
 of the channel $4 \times 4^* \rightarrow 1$
 is expected by the most attractive channel hypothesis.
The confinement of unbroken $SU(4)$
 and a hierarchy between these two condensation scales are expected.
Although
 the expectation of the gauge symmetry breaking of $SU(5) \rightarrow SU(4)$
 may not be true in the supersymmetric theory,
 we construct the effective theory postulating the confinement of $SU(4)$,
 and justify the expectation by the result.
Namely,
 if the vacuum expectation values of the effective fields,
 which are not $SU(5)$-singlet,
 support the gauge symmetry breaking of $SU(5) \rightarrow SU(4)$,
 we can consistently accept this expectation.

In the next section
 we introduce three effective fields, two of which are $SU(5)$-variant.
Since this theory has no flat direction,
 all the effective fields are the massive fields
 except for the Nambu-Goldstone particles.
We assume that global symmetry,
 $U(1)_R$ (R-symmetry) and $U(1)_A$ (chiral symmetry),
 are completely broken,
 and do not apply the 't Hooft anomaly matching condition\cite{matching}.
Is is consistently justified by the finial result.

In section \ref{sec:potential}
 the superpotential is uniquely determined using the method by Seiberg et al.
The non-trivial K\"ahler potential is introduced
 to incorporate the effect of the classical scalar potential,
 by which the diverge vacuum expectation values are forbidden.
The K\"ahler potential is not uniquely determined by the symmetry,
 but we assume most simple form
 by which the quantum scalar potential coincides with the classical one
 in the limit of $\Lambda \rightarrow 0$,
 where $\Lambda$ is the dynamical scale of $SU(5)$ gauge theory.

In section \ref{sec:vacuum}
 the vacuum and particle spectrum are examined.
We find unique vacuum
 with finite vacuum expectation value of effective fields,
 where both two $U(1)$ global symmetry and supersymmetry
 are spontaneously broken.
We show that two Nambu-Goldstone bosons appear
 corresponding with the spontaneous breaking
 of these two global $U(1)$ symmetry,
 one Nambu-Goldstone fermion appears
 corresponding with the breaking of supersymmetry,
 and all the other fields are massive.
All the assumption on the symmetry breaking
 are consistently justified by the result.

In the last section
 we discuss the non-trivial K\"ahler potential, and conclude this paper.

Before closing this section, we settle the notation.
The metric we use is $(1, -1, -1, -1)$,
 and the $\sigma$-matrixes for the two component spinor are
 $(\sigma_\mu)_{\alpha\dot\beta} = (1,\tau^i)$
 and $({\bar\sigma}_\mu)_{\dot\alpha\beta} = (1,-\tau^i)$,
 where $\tau^i$ is the Pauli matrix.
The convention on the contraction of the index of two component spinor is
\begin{equation}
 \theta\theta = \theta^{\dot\alpha} \theta_{\dot\alpha},
\quad
 {\bar\theta}{\bar\theta} = {\bar\theta}^{\alpha} {\bar\theta}_{\alpha},
\end{equation}
 with
 $\theta^{\dot\alpha} = \epsilon^{\dot\alpha\dot\beta} \theta_{\dot\beta}$
 and
 ${\bar\theta}^{\alpha} = \epsilon^{\alpha\beta} {\bar\theta}_{\beta}$,
 where $\epsilon^{\dot\alpha\dot\beta} = \epsilon_{\dot\alpha\dot\beta}$
 and $\epsilon^{\alpha\beta} = \epsilon_{\alpha\beta}$.
The integration over the spinors is defined as
\begin{equation}
 \int d^2 \theta \ \theta^2 = 1,
\quad
 \int d^2 {\bar\theta} \ {\bar \theta}^2 = 1.
\end{equation}
The correspondence
 between the standard notation by Wess and Bagger\cite{wess-bagger}
 and ours is given in appendix \ref{app:notation}.

\section{Effective fields}
\label{sec:fields}

We first summarize the classical properties
 of the supersymmetric $SU(5)$ gauge theory
 with chiral superfields in the $5^*$ and $10$ representations.
The Lagrangian is written down as
\begin{eqnarray}
 {\cal L}
 &=& {1 \over {8g^2}}
      \left\{ \int d^2 \theta
              \ {\rm tr}\left( W^{\dot\alpha} W_{\dot\alpha} \right)
            + \int d^2 \bar\theta
              \ {\rm tr}\left( {\bar W}^{\alpha} {\bar W}_{\alpha} \right)
      \right\}
\nonumber\\
&&
    - \int d^2 \theta d^2 \bar\theta
      \left\{ \Phi^{\dag} e^{2V^T} \Phi
            + {\rm tr}\left( \Omega^{\dag} e^{-2V} \Omega e^{-2V^T} \right)
      \right\},
\end{eqnarray}
 where $g$ denotes the gauge coupling constant,
 $W^{\dot\alpha}$ and ${\bar W}^\alpha$ are the chiral superfields
 of the $SU(5)$ gauge field strength composed by the vector superfield $V$,
 and $\Phi$ and $\Omega$ denote the chiral superfields of matters
 in the $5^*$ and $10$ representations, respectively.
At the classical level
 this theory has three global $U(1)$ symmetry
\begin{eqnarray}
 U(1)_X &:& \left\{
             \begin{array}{l}
              W_{\dot\beta}(y,\theta)
               \rightarrow e^{-i\alpha} W_{\dot\beta}(y,\theta e^{i\alpha}),\\
              \Phi(y,\theta) \rightarrow \Phi(y,\theta e^{i\alpha}),\\
              \Omega(y,\theta) \rightarrow \Omega(y,\theta e^{i\alpha}),
             \end{array}
            \right.
\\
 U(1)_{5^*} &:& \Phi(y,\theta) \rightarrow e^{i\beta} \Phi(y,\theta),
\\
 U(1)_{10} &:& \Omega(y,\theta) \rightarrow e^{i\gamma} \Omega(y,\theta),
\end{eqnarray}
 but only two combinations of
 $Q_R \equiv Q_X + 9 Q_{5^*} - Q_{10}$ and $Q_A \equiv 3 Q_{5^*} - Q_{10}$
 are anomaly free,
 where $Q$'s denote the charge of $U(1)$ rotations.
 \footnote{The $\Theta$-term is unphysical in this theory,
           because there is $U(1)$ symmetry
           which is explicitly broken only by the gauge anomaly.}
The $U(1)$ symmetry
 defined by the charges $Q_R$ and $Q_A$ are called $U(1)_R$ and $U(1)_A$.
The charges of the fields
 $W^{\dot\alpha}$, $\Phi$ and $\Omega$ are as follows.
\begin{equation}
\begin{array}{ccc}
                            & ~~~Q_R~~~   & ~~~Q_A~~~ \\
 ~~~W^{\dot\alpha}~~~       &    -1       &    0      \\
 ~~~\Phi~~~                 &    9        &    3      \\
 ~~~\Omega~~~               &    -1       &    -1     \\
\end{array}
\end{equation}

The classical scaler potential
 comes from the D component of the vector superfields,
\begin{equation}
 V_D = {1 \over {2g^2}} D^a D^a
\label{classical-in-original}
\end{equation}
 with
\begin{eqnarray}
 D^a &=&
  g^2 \left\{ A_\Phi^{\dag} \left(-T^a\right)^T A_\Phi
            + {\rm tr} \left( A_\Omega^{\dag} T^a A_\Omega \right)
            + {\rm tr} \left( A_\Omega^{\dag} A_\Omega (T^a)^T \right)
      \right\}
\nonumber\\
     &=&
  g^2 \left\{ A_\Phi^{\dag} T_{5^*}^a A_\Phi
            + A_\Omega^{\dag} T_{10}^a A_\Omega
      \right\},
\end{eqnarray}
 where $A_\Phi$, $A_\Omega$ are the scalar components
 of the chiral superfields $\Phi$ and $\Omega$,
 and $T^a$, $T_{5^*}^a$ and $T_{10}^a$ denote the generators of $SU(5)$
 for the $5$, $5^*$ and $10$ representations, respectively.
It is well known that
 $A_\Phi=0$ and $A_\Omega=0$ is the unique solution
 of the stationary condition of this potential,
 and classical vacuum is supersymmetric.

It is remarkable that
 no gauge invariant superpotential can be written down.
Since all the gauge invariant holomorphic polynomial
 composed by the chiral superfields $\Phi$ and $\Omega$ vanish,
 we can not consider non-trivial superpotential
 even the non-renormalizable one.
This fact seems to mean that
 the generation of the superpotential by the quantum effect
 can not be expected.
On the other hand,
 the explicit non-perturbative calculation suggests
 that the supersymmetry is spontaneously broken\cite{instanton1,instanton2}.
Therefore,
 it is likely to imagine that
 the superfields $\Phi$ and $\Omega$ are not the good physical fields
 below the scale $\Lambda$ (dynamical scale of $SU(5)$ gauge theory),
 and the low energy effective theory with different fields (composite fields)
 should be considered.
The superpotential in the effective theory will lift up the vacuum,
 and supersymmetry is spontaneously broken.
 \footnote{Of course, we can also imagine that
           the complicated K\"ahler potential
           triggers the spontaneous breaking of supersymmetry
           in both the original and effective theories\cite{effective}.}

Now we consider what effective fields are appropriate in this theory.
Since we do not know the symmetry of the exact vacuum
 unlike in the case of supersymmetric QCD,
 we must assume it.
Namely,
 if we assume both or one of $U(1)_R$ or $U(1)_A$ symmetry
 is the symmetry of the vacuum,
 the effective fields must satisfy
 't Hooft anomaly matching condition\cite{matching}.
It is shown in ref. \cite{effective} that
 if both $U(1)_R$ and $U(1)_A$ symmetry
 is assumed as the symmetry of the vacuum,
 many effective fields with complicated quantum numbers have to be introduced.
Such effective fields describe the particles
 which couple with the complicated high dimensional composite operators
 in the original theory.
This situation is implausible,
 because the dynamics will make light composite particles
 which couple with low dimensional operators.
In this paper we assume that
 both $U(1)_R$ and $U(1)_A$ symmetry are spontaneously broken,
 and there is no massless fermions except for the Nambu-Goldstone fermion
 due to the spontaneous breaking of supersymmetry.
Therefore, 't Hooft anomaly matching condition is not imposed.
This assumption must be justified by the result of the analysis.

As explained in the previous section,
 the confinement of $SU(4)$, a subgroup of $SU(5)$,
 is assumed rather than the confinement of $SU(5)$ itself.
This is also the assumption
 which must be justified by the result of the analysis.
The guiding principle
 to introduce the $SU(5)$ variant effective fields is as follows.

We introduce only the effective fields
 which couple with the Lorentz invariant bi-linear operators
 composed by the three fields $\Phi$, $\Omega$ and $W$ in the original theory.
In addition,
 we assume that the effective fields are
 in the smallest representations of $SU(5)$ in each bi-linear combinations.
Namely, we consider the following effective fields.
\begin{equation}
 \begin{array}{ll}
  X^i \sim \epsilon^{ijklm} \Omega_{jk} \Omega_{lm} &
  \qquad 10 \times 10 \rightarrow 5^* \\
  Y_i \sim \Phi^j \Omega_{ji} &
  \qquad 5^* \times 10 \rightarrow 5 \\
  S \sim {\rm tr} \left( W^{\dot\alpha} W_{\dot\alpha} \right) &
  \qquad 24 \times 24 \rightarrow 1
 \end{array}
\end{equation}
The operator corresponding $5^* \times 5^* \rightarrow 10^*$ vanishes,
 since the superfields commute each other.
Furthermore,
 since we assume the confinement of $SU(4)$,
 only the $SU(4)$-singlet parts of each effective fields
 are introduced as the effective fields.
We introduce three fields
\begin{equation}
 X \equiv X^{i=5}, \quad Y \equiv Y_{i=5}, \quad S,
\end{equation}
 as the effective fields below the scale $\Lambda$.

This guiding principle is supported by the following arguments.
The classical scalar potential eq.(\ref{classical-in-original})
 can be written like
\begin{eqnarray}
 V_D &=& {{g^2} \over 2}
       \left[
        \left( A_\Omega^{\dag} T^a_{10} A_\Omega \right)
        \left( A_\Omega^{\dag} T^a_{10} A_\Omega \right) + \cdots
       \right]
\nonumber\\
     &=& {{g^2} \over 2}
       \left[ -\lambda(10,10,5^*){1 \over {4!}}
               \big( \epsilon \ A_\Omega A_\Omega \big)^i
               \big( \epsilon \ A_\Omega^{\dag} A_\Omega^{\dag} \big)_i
              -\lambda(10,10,50)
               \left| \left( A_\Omega A_\Omega \right)_{50} \right|^2
              + \cdots
       \right],
\end{eqnarray}
 where
\begin{eqnarray}
 \lambda(10,10,5^*)
  &=& {1 \over 2} \left\{ C_2(10)+C_2(10)-C_2(5^*) \right\} = {{12} \over 5},
\\
 \lambda(10,10,50)
  &=& {1 \over 2} \left\{ C_2(10)+C_2(10)-C_2(50) \right\} = -{3 \over 5},
\end{eqnarray} 
 and $C_2(r)$ denotes the coefficient of the second Casimir invariant
 of the representation $r$ of $SU(5)$.
 \footnote{The operator correspond to the channel
           $10 \times 10 \rightarrow 45$ vanishes
           because of the Bose statistics of the superfield.}
The method of the auxiliary field
 can be used to introduce the effective fields.
\begin{eqnarray}
 V_D \rightarrow
  &-& {{g^2} \over 2}\lambda(10,10,5^*){1 \over {4!}}
   \big( \epsilon \ A_\Omega A_\Omega \big)^i
   \big( \epsilon \ A_\Omega^{\dag} A_\Omega^{\dag} \big)_i
\nonumber\\
  &&+ {1 \over 2}\lambda(10,10,5^*){1 \over {4!}}
   \left\{
    g \big( \epsilon \ A_\Omega A_\Omega \big)^i - A_X^i
   \right\}
   \left\{
    g \big( \epsilon \ A_\Omega^{\dag} A_\Omega^{\dag} \big)_i - A_{Xi}^{\dag}
   \right\}
\nonumber\\
  &&+ \cdots
\nonumber\\
 =&& {1 \over 2}\lambda(10,10,5^*){1 \over {4!}} A_X^i A_{Xi}^{\dag}
\nonumber\\
  &&- {g \over 2}\lambda(10,10,5^*){1 \over {4!}}
      \left\{
        \big( \epsilon \ A_\Omega A_\Omega \big)^i A_{Xi}^{\dag}
       + A_X^i \big( \epsilon \ A_\Omega^{\dag} A_\Omega^{\dag} \big)_i
      \right\}
\nonumber\\
  &&+ \cdots,
\end{eqnarray}
 where $A_X^i$ denotes the scalar component of the effective field $X^i$.
This result shows that
 if the coefficient $\lambda$ is positive,
 the classical squared mass of the effective field becomes positive,
 and it is worth considering.
The effective field in the $50$ representation can not be considered,
 since $\lambda(10,10,50) < 0$, and its classical squared mass is negative.
The same arguments are true
 for the effective fields composed by $\Phi$ and $\Omega$.
The effective field $Y_i$ is worth considering,
 since $\lambda(5,10,5) = {9 \over 5} >0$,
 but the effective field in the $45$ representation can not be considered,
 since $\lambda(5,10,45) = -{1 \over 5} <0$.

From this argument
 we obtain the classical scalar potential
 written by the effective fields $A_X$ and $A_Y$.
\begin{equation}
 V_{\rm classical}
  = \lambda_X A_{Xi}^{\dag} A_X^i + \lambda_Y A_Y^{\dag i} A_{Yi}
  \rightarrow \lambda_X |A_X|^2 + \lambda_Y |A_Y|^2,
\label{classical-in-effective}
\end{equation}
 where $\lambda_X \equiv {1 \over 2} {1 \over {4!}} \lambda(10,10,5^*)$
 and $\lambda_Y \equiv {1 \over 5} \lambda(5,10,5)$.
The normalization of the effective fields $X$ and $Y$
 is determined in this arguments.
\begin{equation}
  X \equiv g \ \epsilon^{5jklm} \Omega_{jk} \Omega_{lm}.
\quad
  Y \equiv g \ \Phi^j \Omega_{j5}
\end{equation}
The normalization of the effective field $S$
 is determined in the next section.

\section{Superpotential and K\"ahler potential}
\label{sec:potential}

Three effective chiral superfields $X$, $Y$ and $S$ with charges
\begin{equation}
\begin{array}{ccc}
               & ~~~Q_R~~~   & ~~~Q_A~~~ \\
 ~~~X~~~       &    -2       &    -2     \\
 ~~~Y~~~       &    8        &    2      \\
 ~~~S~~~       &    -2       &    0      \\
\end{array}
\end{equation}
 are introduced in the previous section.
We construct the effective theory using these fields
 which is effective below the scale $\Lambda$.
Since the fields $X$ and $Y$
 are not covariant under the $SU(5)$ transformation,
 the $SU(5)$ invariant Lagrangian can not be constructed.
Imagine that the theory
 in which only the heavy $SU(5)/SU(4)$ gauge bosons
 ($SU(5)/SU(4)$ gauge bosons and would-be Nambu-Goldstone bosons)
 are integrated out,
 but Higgs is not integrated out.
The theory we want to construct
 is neither the ``linear $\sigma$-model'' nor ``non-linear $\sigma$-model''
 for $SU(5)$ gauge symmetry breaking,
 but ``linear $\sigma$-model'' for the breaking of the global symmetry
 $U(1)_R$, $U(1)_A$ and supersymmetry.
Therefore,
 the invariance under all anomaly-free global symmetry is postulated.

The superpotential is uniquely determined
 using the method by Seiberg et al.\cite{seiberg1,seiberg2,seiberg3}.
The product $XYS^3$ is the unique independent holomorphic product
 which is invariant under $U(1)_R$ and $U(1)_A$ transformation,
 and $S$ is the unique independent holomorphic quantity
 which is invariant under $U(1)_A$ transformation with $U(1)_R$-charge $-2$.
 \footnote{The superpotential must have $U(1)_R$-charge $-2$ in our notation}
Therefore,
 the general form of the superpotential is
\begin{equation}
 W = S f \left( {{\Lambda^{13}} \over {XYS^3}} \right)
\end{equation}
 with a general holomorphic function $f$.
Note that the power of $\Lambda$, $13$,
 which comes from the dimensional analysis,
 is just the coefficient of the 1-loop $\beta$-function
 of the $SU(5)$ gauge coupling.
In the weak coupling limit, $\Lambda \rightarrow 0$,
 this superpotential must coincide with the gauge kinetic term
 in the perturbatively-calculated Wilsonian action of the original theory
\begin{equation}
 {\cal L}_{\rm gauge}\Big|_{\Lambda \rightarrow 0}
  = - {1 \over {64\pi^2}} \ln \Lambda^{13}\
      {\rm tr} \left( W^{\dot\alpha}  W_{\dot\alpha} \right) + {\rm h.c.}.
\end{equation}
Therefore,
\begin{equation}
 W = - {1 \over {64\pi^2}}
       \ S \ \ln \left( {{\Lambda^{13}} \over {XYS^3}} \right)
     + S {\tilde f} \left( {{\Lambda^{13}} \over {XYS^3}} \right),
\end{equation}
 where $\tilde f$ is a holomorphic function
 with $\lim_{z \rightarrow 0} {\tilde f}(z) = 0$.
Here we take the normalization
 $S \equiv {\rm tr}(W^{\dot\alpha} W_{\dot\alpha})$.
Moreover,
 if we assume that
 the massless degrees of freedom are only the Nambu-Goldstone particles,
 and all of them are described by the effective fields already introduced,
 the function $\tilde f$ should not have the singularities,
 and it is a constant.
The constant is absorbed to the redefinition of $\Lambda$.
Thus, we obtain
\begin{equation}
 W = - {1 \over {64\pi^2}}
       \ S \ \ln \left( {{\Lambda^{13}} \over {XYS^3}} \right).
\label{superpotential}
\end{equation}
This is the unique superpotential within our postulations.

We can obtain a scalar potential
 from the superpotential of eq.(\ref{superpotential})
 assuming a naive K\"ahler potential
\begin{equation}
 K_{\rm naive} = {1 \over {\Lambda^2}} X^{\dag} X
               + {1 \over {\Lambda^2}} Y^{\dag} Y
               + {1 \over {\Lambda^4}} S^{\dag} S,
\end{equation}
 where the normalization comes from that
 the effective fields $X$ and $Y$, and $S$
 have dimension $2$ and $3$, respectively.
But the solution of the stationary condition of the scalar potential
 is $\langle A_S \rangle \rightarrow 0$
 and $\langle A_X \rangle, \langle A_Y \rangle \rightarrow \infty$
 with $\ln (\Lambda^{13}/\langle A_X \rangle \langle A_Y \rangle
                         \langle A_S \rangle^3) = 3$
 (supersymmetric vacuum).
This solution is not acceptable,
 because $\langle A_X \rangle$ and $\langle A_Y \rangle$
 should not become infinity by virtue of the classical potential
 of eq.(\ref{classical-in-effective}).
The effect of the classical potential must be included.
It is not the supersymmetric treatment
 to simply add the classical potential to the quantum potential,
 because the classical potential is the explicit soft breaking term
 of supersymmetry.
It is also impossible
 to include the classical effect as a constraint in the superpotential
 using the Lagrange multiplier like in the supersymmetric QCD with $N_f=N_c$,
 because this theory has no flat direction in the scalar potential.
The remaining possibility is to consider the non-trivial K\"ahler potential.

The non-trivial K\"ahler potential of the form
\begin{equation}
 K(X^{\dag}X,Y^{\dag}Y,S^{\dag}S)
  = K_X(X^{\dag}X) + K_Y(Y^{\dag}Y) + K_S(S^{\dag}S)
\label{kaehler}
\end{equation}
 modifies the equation of motion
 of the auxiliary fields of each effective fields as
\begin{eqnarray}
 F_X^{\dag} = - \left[ {{\partial W} \over {\partial X}} \right] \Bigg/
                \left[ {\partial \over {\partial (X^{\dag}X)}}
                 \left(
                  (X^{\dag}X){{\partial K_X} \over {\partial (X^{\dag}X)}}
                 \right)
                \right],
\end{eqnarray}
 and so on, where $[ \ ]$ denotes to take the scalar component.
The scalar potential is given by
\begin{equation}
 V = - \left[ {{\partial W} \over {\partial X}} \right]^{\dag} F_X^{\dag}
     - \left[ {{\partial W} \over {\partial Y}} \right]^{\dag} F_Y^{\dag}
     - \left[ {{\partial W} \over {\partial S}} \right]^{\dag} F_S^{\dag}.
\label{potential-from-F}
\end{equation}

We consider the K\"ahler potential
\begin{eqnarray}
 K_X(X^{\dag}X) &=& {1 \over {\Lambda^2}} f(X^{\dag}X)_{C_X/\Lambda},
\\
 K_Y(Y^{\dag}Y) &=& {1 \over {\Lambda^2}} f(Y^{\dag}Y)_{C_Y/\Lambda},
\\
 K_S(S^{\dag}S) &=& {1 \over {\Lambda^4}} S^{\dag}S,
\end{eqnarray}
 with two real parameters $C_X$ and $C_Y$,
 where a function $f(z)_a$ is defined by
\begin{equation}
 f(z)_a \equiv \sum_{n=0}^\infty (-1)^n {{a^{2n}z^{2n+1}} \over {(2n+1)^2}}
        = z \ F \left(1, {1 \over 2}, {1 \over 2};
                      {3 \over 2}, {3 \over 2}; -a^2 z^2 \right),
\end{equation}
 and it satisfies
\begin{equation}
 {d \over {dz}} \left( z \ {{d f(z)_a} \over {dz}} \right)
  = {1 \over {1+a^2 z^2}}.
\end{equation}
The function $F$ is the generalized hypergeometric function.
Note that the naive K\"ahler potential
 is contained in $K_X$ and $K_Y$ as the first term of the expansion.

The scalar potential is obtained as
\begin{eqnarray}
 V &=& {{\Lambda^4} \over {(64 \pi^2)^2}}
       \left|
        \ln \left( {{A_X A_Y A_S^3} \over {\Lambda^{13}}} \right) + 3
       \right|^2
    +  {{\Lambda^2} \over {(64 \pi^2)^2}}
       \left(
        {{|A_S|^2} \over {|A_X|^2}} + {{|A_S|^2} \over {|A_Y|^2}}
       \right)
\nonumber\\
   &&+ {{C_X^2} \over {(64 \pi^2)^2}} {{|A_S|^2} \over {\Lambda^6}} |A_X|^2
     + {{C_Y^2} \over {(64 \pi^2)^2}} {{|A_S|^2} \over {\Lambda^6}} |A_Y|^2.
\label{potential}
\end{eqnarray}
The last two terms are the contribution of the non-trivial K\"ahler potential.

The two parameters $C_X$ and $C_Y$ are determined
 so that the potential of eq.(\ref{potential})
 coincides with the classical one, eq.(\ref{classical-in-effective}),
 in $\Lambda \rightarrow 0$ limit.
The first two terms of the potential simply vanish in this limit,
 but the last two terms seem to be singular.
As taking the limit,
 the vacuum expectation value of the effective field
 takes the place of its dynamical degree of freedom,
 and the effective field decouples.
We assume that
 the effective field $S$ firstly decouples
 because of its largest vacuum expectation value,
 though this should be justified by the result.
The vacuum expectation value of $S$ is proportional to $\Lambda^3$,
 and the coefficient $r$ is independent of $\Lambda$,
 but it depends on $C_X$ and $C_Y$.
Therefore,
 we can determine these two parameters by the condition
\begin{equation}
 {{C_X^2} \over {(64 \pi^2)^2}} \ r(C_X,C_Y)^2 = \lambda_X,
\quad
 {{C_Y^2} \over {(64 \pi^2)^2}} \ r(C_X,C_Y)^2 = \lambda_Y.
\end{equation}
In practice,
 we replace the parameters $C_X$ and $C_Y$
 by $\lambda_X$, $\lambda_Y$ and $r$,
 and iteratively solve the stationary condition of the scalar potential
 changing the value of $r$
 until finding the solution $\langle A_S \rangle = r \Lambda^3$.

\section{Vacuum and mass spectrum}
\label{sec:vacuum}

Now we solve the stationary condition of the potential of eq.(\ref{potential}).
By using the phase rotation of $U(1)_R$ and $U(1)_A$,
 the vacuum expectation values of $A_X$ and $A_Y$
 can be taken as real positive.
The vacuum expectation value of $A_S$ can have the imaginary part,
 but it is dynamically set to zero.
Substituting $A_S = |A_S| e^{i\theta_S/\Lambda}$ into the potential,
 we obtain the potential for $\theta_S$ as
\begin{equation}
 V_{\theta_S} = {{9\Lambda^2} \over {(64 \pi^2)^2}} \theta_S^2.
\end{equation}
Therefore,
 $\langle \theta_S \rangle = 0$,
 and the vacuum expectation value of $A_S$
 also can be taken as real positive.

The stationary conditions
 on the three real positive valuables $A_X$, $A_Y$ and $A_S$ are
\begin{equation}
 {{\Lambda^4} \over {(64 \pi^2)^2}}
  \left( \ln {{A_X A_Y A_S^3} \over {\Lambda^{13}}} + 3 \right)
 - {{\Lambda^2} \over {(64 \pi^2)^2}} {{A_S^2} \over {A_X^2}}
 + {{\lambda_X} \over {r^2}} {{A_S^2} \over {\Lambda^6}} A_X^2 = 0,
\label{stationary-X}
\end{equation}
\begin{equation}
 {{\Lambda^4} \over {(64 \pi^2)^2}}
  \left( \ln {{A_X A_Y A_S^3} \over {\Lambda^{13}}} + 3 \right)
 - {{\Lambda^2} \over {(64 \pi^2)^2}} {{A_S^2} \over {A_Y^2}}
 + {{\lambda_Y} \over {r^2}} {{A_S^2} \over {\Lambda^6}} A_Y^2 = 0,
\label{stationary-Y}
\end{equation}
\begin{equation}
 {{3\Lambda^4} \over {(64 \pi^2)^2}}
  \left( \ln {{A_X A_Y A_S^3} \over {\Lambda^{13}}} + 3 \right)
 + {{\Lambda^2} \over {(64 \pi^2)^2}}
  \left( {{A_S^2} \over {A_X^2}} + {{A_S^2} \over {A_Y^2}} \right)
 + {{\lambda_X} \over {r^2}} {{A_S^2} \over {\Lambda^6}} A_X^2
 + {{\lambda_X} \over {r^2}} {{A_S^2} \over {\Lambda^6}} A_Y^2 = 0,
\label{stationary-S}
\end{equation}
 respectively,
 where two parameters $C_X$ and $C_Y$
 are replaced by the two known parameters $\lambda_X$ and $\lambda_Y$,
 and an unknown parameter $r$.

Although it is difficult
 to get the complete analytical solution of these conditions,
 an analytical relation
\begin{equation}
 \langle A_Y \rangle^2
  = {{2 \langle A_X \rangle^2}
     \over
     {(64 \pi^2)^2 (\lambda_X / r^2) \langle A_X \rangle^4 / \Lambda^8 - 3}}
\end{equation}
 is obtained.
We substitute this relation
 into eqs.(\ref{stationary-X}) and (\ref{stationary-Y}),
 and numerically solve them.
We find the solution
\begin{equation}
 \langle A_X \rangle \simeq (0.17)^2,
\quad
 \langle A_Y \rangle \simeq (0.11)^2,
\quad
 \langle A_S \rangle \simeq (0.31)^3,
\label{solution}
\end{equation}
 in unit of $\Lambda$ with $r=0.03$.
Note that consistently $r=0.03 \simeq \langle A_S \rangle \simeq 0.031$,
 and the vacuum expectation value of the effective field $A_S$
 is the largest one, namely,
 $\langle A_S \rangle^{1/3} > \langle A_X \rangle^{1/2}
                            > \langle A_Y \rangle^{1/2}$.

This solution
 is consistent with the assumption of breaking $SU(5) \rightarrow SU(4)$,
 since the effective field $X$,
 which is a component of the $SU(5)$-variant effective field
 in the $5^*$ representation,
 obtains the vacuum expectation value.
It can be considered that
 the vacuum expectation value of the effective field $Y$
 is caused by the dynamics of the effective $SU(4)$ gauge theory
 below the scale of the breaking of $SU(5)$
 triggered by $\langle X \rangle \neq 0$.
The assumption on the breaking 
 of the global $U(1)_R \times U(1)_A$ symmetry is also confirmed.
Since the vacuum expectation value of the effective filed $S$
 means the gaugino pair condensation,
 the spontaneous breaking of supersymmetry is expected
 through Konishi anomaly\cite{konishi}.
In fact,
 the vacuum energy density is not zero,
 $V_{\rm vacuum} \simeq (0.16)^4$ in unit of $\Lambda$.
The vacuum energy density
 is the order parameter of supersymmetry breaking
 with absolute normalization.
Taking $\langle A_X \rangle$
 as the order parameter of gauge symmetry breaking,
 we find that both supersymmetry and gauge symmetry
 are spontaneously broken at almost the same scale.

The mass spectrum of the effective fields can be explicitly calculated.
On boson fields,
 it is convenient to consider the non-linear realization
 of the global $U(1)_R \times U(1)_A$ symmetry
\begin{equation}
 A_X = \Lambda \ \phi_X e^{i\theta_X/\Lambda},
\quad
 A_Y = \Lambda \ \phi_Y e^{i\theta_Y/\Lambda},
\quad
 A_S = \Lambda^2 \ \phi_S e^{i\theta_S/\Lambda},
\end{equation}
 where $\phi_{X,Y,S}$ and $\theta_{X,Y,S}$ are the real scalar fields
 with dimension one.
By substituting this expression
 to the scaler potential of eq.(\ref{potential}),
 we obtain
\begin{eqnarray}
 V &=& {{\Lambda^4} \over {(64 \pi^2)^2}}
       \left\{
        \left( \ln {{\phi_X \phi_Y \phi_S^3} \over {\Lambda^5}} + 3 \right)^2
        + {1 \over {\Lambda^2}}
          \left( \theta_X + \theta_Y + 3 \theta_S \right)^2
       \right\}
\nonumber\\
   &&+ {{\Lambda^4} \over {(64 \pi^2)^2}}
       \left(
        {{\phi_S^2} \over {\phi_X^2}} + {{\phi_S^2} \over {\phi_Y^2}}
       \right)
     + {{\lambda_X} \over {r^2}} \phi_S^2 \phi_X^2
     + {{\lambda_Y} \over {r^2}} \phi_S^2 \phi_Y^2.
\label{potential2}
\end{eqnarray}
The mass matrix for the fields $\theta_{X,Y,S}$ is given by
\begin{equation}
 {\cal L}_{\rm mass}^\theta
  = - {1 \over 2}
    \left(
    \begin{array}{ccc}
     \theta_X & \theta_Y & \theta_S
    \end{array}
    \right)
    M_\theta^2
    \left(
    \begin{array}{c}
     \theta_X \\ \theta_Y \\ \theta_S
    \end{array}
    \right),
\quad
 M_\theta^2 = 
 {{2\Lambda^2} \over {(64 \pi^2)^2}}
 \left(
 \begin{array}{ccc}
  1 & 1 & 3 \\
  1 & 1 & 3 \\
  3 & 3 & 9 
 \end{array}
 \right).
\end{equation}
Two of three eigenvalues are zero
 which are corresponding to the Nambu-Goldstone bosons
 of $U(1)_R$ and $U(1)_A$ breaking,
 and remaining eigenvalue is
 $m_\theta^2 = 22\Lambda^2/(64 \pi^2)^2 \simeq (0.0074 \Lambda)^2$.
The smallness of this value can be understood by considering that
 it is corresponding to the mass of the pseudo-Nambu-Goldstone boson
 due to the anomalous global $U(1)$ symmetry breaking.

The mass matrix for the fields $\phi_{X,Y,S}$
 can be obtained by differentiating the potential of eq.(\ref{potential2}).
\begin{equation}
 {\cal L}_{\rm mass}^A
  = - {1 \over 2}
    \left(
    \begin{array}{ccc}
     \phi_X & \phi_Y & \phi_S
    \end{array}
    \right)
    M_A^2
    \left(
    \begin{array}{c}
     \phi_X \\ \phi_Y \\ \phi_S
    \end{array}
    \right),
\end{equation}
 where $M_A^2$ is given by
\begin{equation}
 {1 \over {(64 \pi^2)^2}}
 \left(
 \begin{array}{ccc}
  \begin{array}{l}
  2 \Bigg\{
      {1 \over {A_X^2}}
    + {{3 A_S^2} \over {A_X^4}}
  \\
    + (64 \pi^2)^2 {{\lambda_X} \over {r^2}} A_S^2
  \\
    - {{\ln (A_X A_Y A_S^3) + 3} \over {A_X^2}}
    \Bigg\}
  \end{array}
 &
  {1 \over {A_X A_Y}}
 &
  \begin{array}{l}
     {{3} \over {A_S A_X}}
   - {{2 A_S} \over {A_X^3}}
  \\
   + (64 \pi^2)^2 {{2 \lambda_X} \over {r^2}} A_S A_X
  \end{array}
 \\
  {1 \over {A_X A_Y}}
 &
  \begin{array}{l}
  2 \Bigg\{
      {1 \over {A_Y^2}} + {{3 A_S^2} \over {A_Y^4}}
  \\
    + (64 \pi^2)^2 {{\lambda_Y} \over {r^2}} A_S^2
  \\
    - {{\ln (A_X A_Y A_S^3) + 3} \over {A_Y^2}}
    \Bigg\}
  \end{array}
 &
  \begin{array}{l}
     {3 \over {A_S A_Y}}
   - {{2 A_S} \over {A_Y^3}}
  \\
   + (64 \pi^2)^2 {{2 \lambda_Y} \over {r^2}} A_S A_Y
  \end{array}
 \\
  \begin{array}{l}
     {3 \over {A_S A_X}}
   - {{2 A_S} \over {A_X^3}}
  \\
   + (64 \pi^2)^2 {{2 \lambda_X} \over {r^2}} A_S A_X
  \end{array}
 &
  \begin{array}{l}
     {3 \over {A_S A_Y}}
   - {{2 A_S} \over {A_Y^3}}
  \\
   + (64 \pi^2)^2 {{2 \lambda_Y} \over {r^2}} A_S A_Y
  \end{array}
 &
  \begin{array}{l}
  2 \Bigg\{
      {1 \over {A_X^2}} + {1 \over {A_Y^2}}
  \\
    + (64 \pi^2)^2 \left(
                     {{\lambda_X} \over {r^2}} A_X^2
                   + {{\lambda_Y} \over {r^2}} A_Y^2
                   \right)
  \\
    - {{3 \ln (A_X A_Y A_S^3)} \over {A_S^2}}
    \Bigg\}
  \end{array}
 \end{array}
 \right).
\end{equation}
Here $\Lambda$ is set to unity,
 and $A_X$, $A_Y$ and $A_S$ denote the vacuum expectation values
 of eq.(\ref{solution}).
Though the analytic expression of the mass matrix is very complicated,
 its three eigenvalues can be numerically estimated as
\begin{equation}
 m_A^2 \simeq (0.45)^2, \quad (0.73)^2, \quad (1.5)^2,
\end{equation}
 in unit of $\Lambda$.
No large hierarchy is realized, but all are rather heavy.

There are two contributions
 to the masses of the fermion components,
 $\Lambda \psi_X$, $\Lambda \psi_Y$ and $\Lambda^2 \psi_S$,
 of the effective chiral superfields, $X$, $Y$ and $S$, respectively,
 where all $\psi$'s have dimension $3/2$.
One comes from the superpotential,
 and another comes from the K\"ahler potential.
The superpotential of eq.(\ref{superpotential})
 gives the mass matrix
\begin{equation}
 {\cal L}_{\rm mass}^{W}
 = - {1 \over 2}
    \left(
    \begin{array}{ccc}
     \psi_X^{\dot\alpha} & \psi_Y^{\dot\alpha} & \psi_S^{\dot\alpha}
    \end{array}
    \right)
    M_\psi^W
    \left(
    \begin{array}{c}
     \psi_{X\dot\alpha} \\ \psi_{Y\dot\alpha} \\ \psi_{S\dot\alpha}
    \end{array}
    \right)
   + {\rm h.c.},
\end{equation}
\begin{equation}
 M_\psi^W = 
 {1 \over {64 \pi^2}}
 \left(
 \begin{array}{ccc}
  {{3 \Lambda^4} \over {\langle A_S \rangle}} &
  {{\Lambda^3} \over {\langle A_X \rangle}} &
  {{\Lambda^3} \over {\langle A_Y \rangle}} \\
  {{\Lambda^3} \over {\langle A_X \rangle}} &
  - {{\Lambda^2 \langle A_S \rangle} \over {\langle A_X \rangle^2}} &
  0 \\
  {{\Lambda^3} \over {\langle A_Y \rangle}} &
  0 &
  - {{\Lambda^2 \langle A_S \rangle} \over {\langle A_Y \rangle^2}} \\
 \end{array}
 \right).
\end{equation}

The mass matrix
 emerged from the K\"ahler potential of eq.(\ref{kaehler})
 has diagonal from.
Since the K\"ahler potential for $S$ is naive one,
 there is no contribution to the mass of $\psi_S$.
The contribution to the masses of remaining $\psi_X$ and $\psi_Y$ is
\begin{eqnarray}
 {\cal L}_{\rm mass}^K
 &=& {{\left\langle
        {{\partial^2 K_X} \over {\partial (X^{\dag}X)^2}}
        + {1 \over 2} ( X^{\dag}X )
          {{\partial^3 K_X} \over {\partial (X^{\dag}X)^3}}
       \right\rangle}
      \over
      {\left\langle
        {\partial \over {\partial (X^{\dag}X)}}
        \left( (X^{\dag}X){{\partial K_X} \over {\partial (X^{\dag}X)}} \right)
       \right\rangle}}
     \left\langle {{\partial W} \over {\partial X}} X^{\dag} \right\rangle
     \psi_X^{\dot\alpha} \psi_{X\dot\alpha} + {\rm h.c.}
\nonumber\\
 &&+ {{\left\langle
        {{\partial^2 K_Y} \over {\partial (Y^{\dag}Y)^2}}
        + {1 \over 2} ( Y^{\dag}Y )
          {{\partial^3 K_Y} \over {\partial (Y^{\dag}Y)^3}}
       \right\rangle}
      \over
      {\left\langle
        {\partial \over {\partial (Y^{\dag}Y)}}
        \left( (Y^{\dag}Y){{\partial K_Y} \over {\partial (Y^{\dag}Y)}} \right)
       \right\rangle}}
     \left\langle {{\partial W} \over {\partial Y}} Y^{\dag} \right\rangle
     \psi_Y^{\dot\alpha} \psi_{Y\dot\alpha} + {\rm h.c.},
\end{eqnarray}
 where $\langle \ \rangle$ denotes
 to take the vacuum expectation value of the scalar component.
Since the function $f(z)_a$ satisfies the simple formula
\begin{equation}
 {{d^2 f(z)_a} \over {d z^2}} + {1 \over 2} z {{d^3 f(z)_a} \over {d z^3}}
 = -{{a^2 z} \over {(1+a^2 z^2)^2}},
\end{equation}
 we obtain
\begin{equation}
 {\cal L}_{\rm mass}^K
 = - {1 \over {64 \pi^2}}
     {{C_X^2 \langle A_X \rangle^2 \langle A_S \rangle / \Lambda^6}
      \over
      {1 + C_X^2 \langle A_X \rangle^4 / \Lambda^8}}
     \psi_X^{\dot\alpha} \psi_{X\dot\alpha}
   - {1 \over {64 \pi^2}}
     {{C_Y^2 \langle A_Y \rangle^2 \langle A_S \rangle / \Lambda^6}
      \over
      {1 + C_Y^2 \langle A_Y \rangle^4 / \Lambda^8}}
     \psi_Y^{\dot\alpha} \psi_{Y\dot\alpha}
   + {\rm h.c.}.
\end{equation}
Therefore,
 total mass matrix becomes
\begin{equation}
 M_\psi = {1 \over {64 \pi^2}}
 \left(
  \begin{array}{ccc}
   {{3 \Lambda^4} \over {\langle A_S \rangle}} &
   {{\Lambda^3} \over {\langle A_X \rangle}} &
   {{\Lambda^3} \over {\langle A_Y \rangle}} \\
   {{\Lambda^3} \over {\langle A_X \rangle}} &
   -{{\Lambda^2 \langle A_S \rangle} \over {\langle A_X \rangle^2}}
    {{1-C_X^2 \langle A_X \rangle^4 / \Lambda^8}
     \over
     {1+C_X^2 \langle A_X \rangle^4 / \Lambda^8}} &
   0 \\
   {{\Lambda^3} \over {\langle A_X \rangle}} &
   0 &
   -{{\Lambda^2 \langle A_S \rangle} \over {\langle A_Y \rangle^2}}
    {{1-C_Y^2 \langle A_Y \rangle^4 / \Lambda^8}
     \over
     {1+C_Y^2 \langle A_Y \rangle^4 / \Lambda^8}} \\
  \end{array}
 \right).
\end{equation}
One can easily check that
 this mass matrix has one zero eigenvalue,
 which is corresponding to the Nambu-Goldstone fermion
 of supersymmetry breaking,
 by using the stationary conditions of
 eqs.(\ref{stationary-X}), (\ref{stationary-Y}) and (\ref{stationary-S})
 with $C_{X,Y}^2 = (64 \pi^2)^2 \lambda_{X,Y}/r^2$.
The other two eigenvalues are numerically given by
\begin{equation}
 m_\psi \simeq 0.33, \quad 0.091,
\end{equation}
 in unit of $\Lambda$.
There is no large hierarchy.

\section{Conclusion}
\label{sec:conclusion}

An effective theory of the supersymmetric $SU(5)$ gauge theory
 with chiral superfields in the $5^*$ and $10$ representations
 is constructed within two important postulations.
One important postulation is
 on the symmetry breaking of the gauge and global symmetry.
We postulate
 the spontaneous gauge symmetry breaking of $SU(5) \rightarrow SU(4)$
 and confinement of $SU(4)$.
It is also postulated that
 the global symmetry $U(1)_R \times U(1)_A$ is completely broken.
Basing on this postulation,
 the effective fields which are not $SU(5)$-singlet but singlet
 under the transformation of its subgroup $SU(4)$
 are introduced without imposing 't Hooft anomaly matching condition.
The postulation on these symmetry breaking
 is consistently justified by the result.

Another important postulation is
 to introduce the non-trivial K\"ahler potential
 so that the quantum scalar potential coincides with the classical one
 in the limit of $\Lambda \rightarrow 0$.
It is notable that
 the first term of the expansion of the introduced K\"ahler potential
 is the naive one which gives normal kinetic terms of the component fields.

The K\"ahler potential introduced in this paper
 may be the unique one which satisfies the conditions:
\begin{enumerate}
 \item Coincide with the naive K\"ahler potential
       in the limit of weak field strength,
 \item Scalar potential coincides with the classical one
       in the limit of weak coupling.
\end{enumerate}
We can try to introduce the K\"ahler potential
 by which the classical scalar potential is trivially incorporated
 into the quantum scalar potential.
Such K\"ahler potential must have the form
\begin{equation}
 K(X^{\dag}X,Y^{\dag}Y,S^{\dag}S)
  = K_{XS}(X^{\dag}X,S^{\dag}S) + K_{YS}(Y^{\dag}Y,S^{\dag}S),
\end{equation}
 and the equation of motion
 of the auxiliary fields of each effective fields become
\begin{equation}
 F_X^{\dag}
 = - {{\left[ {{\partial W} \over {\partial X}} \right]
       \left[ {\partial \over {\partial (S^{\dag}S)}}
        \left(
         (S^{\dag}S) {{\partial K_{XS}} \over {\partial (S^{\dag}S)}}
        \right)
       \right]
     - \left[ {{\partial^2 K_{XS}}
              \over
              {\partial (X^{\dag}X) \partial (S^{\dag}S)}}
       \right]
       \left[ {{\partial W} \over {\partial S}} \right]
       \left[ X^{\dag} S \right]}
      \over
      {\left[ {\partial \over {\partial (X^{\dag}X)}}
        \left(
         (X^{\dag}X) {{\partial K_{XS}} \over {\partial (X^{\dag}X)}}
        \right)
       \right]
       \left[ {\partial \over {\partial (S^{\dag}S)}}
        \left(
         (S^{\dag}S) {{\partial K_{XS}} \over {\partial (S^{\dag}S)}}
        \right)
       \right]
     - \left[ {{\partial^2 K_{XS}}
              \over
              {\partial (X^{\dag}X) \partial (S^{\dag}S)}}
       \right]^2
       \left[ X^{\dag}X S^{\dag}S \right]}},
\end{equation}
 and so on.
The potential given by eq.(\ref{potential-from-F})
 becomes extremely complicated one,
 and some undesirable terms,
 which are singular in the limit of $\Lambda \rightarrow 0$
 keeping the dynamical degrees of freedom of $X$ and $Y$ alive,
 will emerge.

The mass spectrum of the effective fields
 which describe composite particles are explicitly calculated.
It is analytically shown that
 the three Nambu-Goldstone particles
 which is corresponding with the spontaneous breaking
 of supersymmetry and $U(1)_R \times U(1)_A$ symmetry appear in the spectrum.
There is no large scale hierarchy in the mass spectrum,
 but we can see that bosons except for the (pseudo-)Nambu-Goldstone bosons
 are heavier than the fermions.

It is expected that
 the method developed in this paper
 is applied to the other (chiral) gauge theories with no flat direction,
 and some new dynamics are found,
 by which the problems of the standard model are solved.

\acknowledgments

We would like to thank Nobuchika Okada for helpful discussions.

\appendix
\section{Notation}
\label{app:notation}

Followings are the correspondence
 between the notation by Wess and Bagger\cite{wess-bagger} and ours.

\newcommand{\th}{\theta}
\newcommand{\thb}{\bar \theta}
\newcommand{\ps}{\psi}
\newcommand{\psb}{\bar \psi}
\newcommand{\ch}{\chi}
\newcommand{\chb}{\bar \chi}
\newcommand{\al}{\alpha}
\newcommand{\ald}{{\dot \alpha}}
\newcommand{\be}{\beta}
\newcommand{\bed}{{\dot \beta}}
\newcommand{\ga}{\gamma}
\newcommand{\gad}{{\dot \gamma}}
\newcommand{\de}{\delta}
\newcommand{\ded}{{\dot \delta}}
\newcommand{\sig}{\sigma}
\newcommand{\sigb}{\bar \sigma}

On the metric and spinors:
\begin{equation}
\eta^{mn} \Bigg|_{W-B} = - g^{\mu\nu}.
\end{equation}
\begin{equation}
\epsilon^{\al\be} \Bigg|_{W-B} = \epsilon^{\al\be},
\qquad
\epsilon_{\al\be} \Bigg|_{W-B} = - \epsilon_{\al\be}.
\end{equation}
\begin{equation}
\left( \sig^m \right)_{\al\bed} \Bigg|_{W-B}
 = - \left( \sig^\mu \right)_{\al\bed},
\qquad
\left( \sigb^m \right)^{\ald\be} \Bigg|_{W-B}
 = - \left( \sigb^\mu \right)^{\ald\be}.
\end{equation}
\begin{equation}
\th^\al \Bigg|_{W-B} = \thb^\al,
\qquad
\thb^\ald \Bigg|_{W-B} = \th^\ald.
\end{equation}
\begin{equation}
\th\th \Bigg|_{W-B} = \thb\thb = \thb^\al \thb_\al,
\qquad
\thb\thb \Bigg|_{W-B} = - \th\th = - \th^\ald \th_\ald.
\end{equation}
\begin{equation}
d^2 \th \Bigg|_{W-B} = d^2 \thb,
\qquad
d^2 \thb \Bigg|_{W-B} = - d^2 \th.
\end{equation}

On the chiral superfields:
\begin{equation}
W_\al (x,\th) \Bigg|_{W-B} = {1 \over 2} {\bar W}_\al (x,\thb),
\qquad
{\bar W}_\ald (x,\thb) \Bigg|_{W-B} = {1 \over 2} W_\ald (x,\th).
\end{equation}
\begin{equation}
\Phi (y,\th) \Bigg|_{W-B} = \Phi^{\dag} (y^{\dag},\thb),
\qquad
\Phi^{\dag} (y^{\dag},\thb) \Bigg|_{W-B} = \Phi (y,\th).
\end{equation}
\begin{equation}
y^m \Bigg|_{W-B} \equiv x^m + i \th \sig^m \thb \Bigg|_{W-B}
 = y^{\dag\mu} \equiv x^\mu - i \thb \sig^\mu \th.
\end{equation}

\end{document}